\documentclass{article}

\usepackage{amssymb,amsmath,times}
\usepackage{icrctc07, here}
\title{Weather induced effects on extensive air
showers observed with the surface detector of the
Pierre Auger Observatory}
\shorttitle{Weather effects on extensive air showers}
\authors{Carla Bleve$^1$, for the Pierre Auger Collaboration$^2$}
\shortauthors{The Auger Collaboration}
\afiliations{
$^1$School of Physics and Astronomy, University of Leeds, Leeds, LS2
9JT, United Kingdom\\ 
$^2$Observatorio Pierre Auger, Av. San Mart\'in Norte 304,
(5613) Malarg\"ue, Mendoza, Argentina}
\email{C.Bleve@leeds.ac.uk}
\abstract{The rate of events measured with the surface detector of the Pierre Auger Observatory is found to
be modulated by the weather conditions. This effect is due to the increasing amount of matter
traversed by a shower as the ground pressure increases and to the inverse proportionality of the
Moli\`ere radius to the air density near ground. Air-shower simulations with different realistic
profiles of the atmosphere support this interpretation of the observed effects.}

\begin{document}
\maketitle
%Begin the section.
\section{Introduction}

The surface detector (SD) of the Auger Southern Observatory, located in
Malarg\"{u}e, Argentina,  is designed for the detection of ultra high energy
cosmic rays through the measurement of the signal induced by the shower
particles reaching the observation level ($\sim$~880 g cm$^{-2}$)  in an array
of  water-Cherenkov  tanks arranged in a triangular grid with 1500 m spacing.\\
The regular data taking of the SD started in January 2004, with the array
continuosly growing from 100 stations up to the current 1200.  To check the
detector stability we monitored the shower counting rate, finding that it is 
modulated by weather effects. This dependence is expected since  changes in the
atmospheric density profile due to weather variations influence the development
of the air shower and in turn the amplitude  of the signal  measured at ground.
As a consequence, a study of the detector stability has to account for the rate
dependence on the atmospheric conditions. 
Moreover, since the SD estimate of the energy of the primary particle is based
on $S(1000)$,  the signal measured at 1000 m from the  shower axis,   we are
interested in the dependence of $S(1000)$ on the atmospheric conditions.
This requires a continuous monitoring of the weather and a good knowledge of
the relationship between $S(1000)$  and the measured weather parameters. The
former is provided by a meteorological station, at the centre of the SD array,
that records the weather parameters every 5 min, allowing the correlation of
the modulation of observed quantities, such as the rate of events, with the
measured ground temperature $T$ and  pressure $P$.

%
%
%%%%%%%%%%%%%%%%%%%%%%%%%%%%%%%%%%%%%%%  
\section{Weather effects on EAS}
The expected effects related to the change of weather conditions are essentially
two:

%First effect
(i) an increase in the ground pressure $P$ corresponds 
to an increased slant depth $X$
and implies that the sho\-wer is older when it reaches the ground level.
%%
%%%%%%%%%%%%%%%%%%%%%%%%%%%%%%%%%%%%%%%%%%%%%%%%%%
\begin{figure}[ht]
  \begin{center}
  \includegraphics[width=0.5\textwidth]{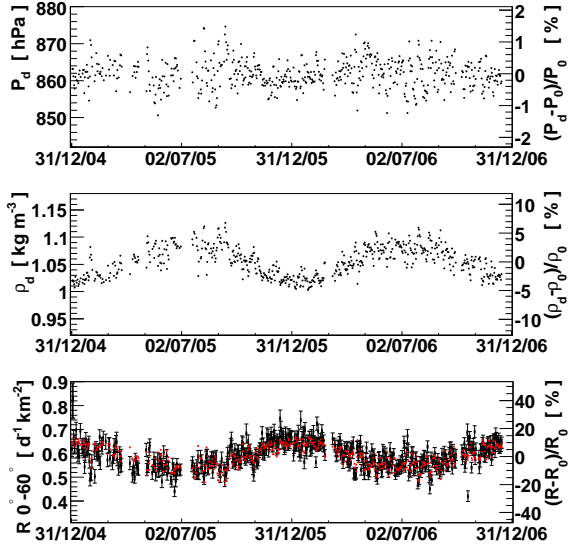}
  \end{center}
  \vspace{-0.2cm}
  \caption{Daily averages of ground pressure (top), density (middle) and event
  rate (bottom, black). Since the pressure is stable, the prominent effect on
  the rate modulation is due to the density (temperature) variation. The red
  points in the bottom plot show the results of the fit.}
  \label{fig:daily}
\end{figure}
%%%%%%%%%%%%%%%%%%%%%%%%%%%%%%%%%%%%%%%%%%%%%%%%%%
%%%%%%%%%%%%%%%%%%%%%%%%%%%%%%%%%%%%%%%%%%%%%%%%%%
\begin{figure}[ht]
  \begin{center}
  \vspace{0.1cm}
  \includegraphics[width=0.5\textwidth]{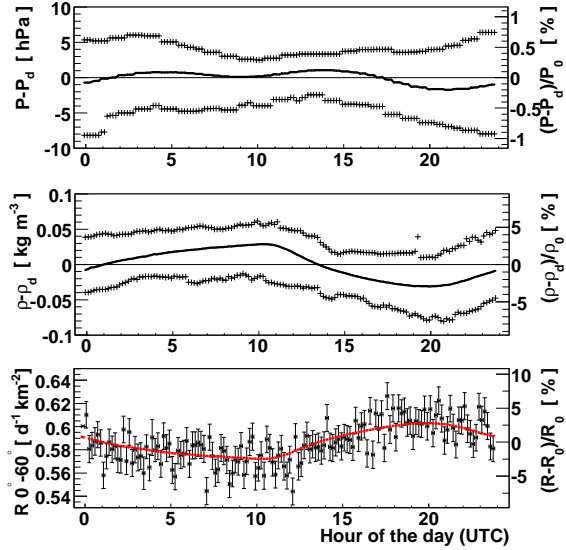}
  \end{center}
  \vspace{-0.2cm}
  \caption{Variation of P (top) and $\rho$ (middle) during the day: the values
  averaged over 2005 and 2006 (lines) are shown together with the maximum
  variation values during the  2 years considered (crosses). Bottom: the result
  of the fit (red) reproduces very well  the average diurnal modulation of the
  measured rate (black). The local time is UTC - 3 h} 
  \label{fig:diurnal}
\end{figure}
%%%%%%%%%%%%%%%%%%%%%%%%%%%%%%%%%%%%%%%%%%%%%%%%%%
%
%
%%%%%%%%%%%%%%%%%%%%%%%%%%%%%%%%%%%%%%%%%%%%%%%%%%%%%%%%%%%%%%%%%%%%%%%%%%%%%%%%%%%%%%%%%%%
%
The longitudinal development of the electromagnetic component of the shower at
1~km from the core can be parameterised as a  Gaisser-Hillas profile,
$N_{em}(E,X)\propto X^{\hat X_m/\Lambda}\exp[(\hat X_m-X)/\Lambda]$,
where $\hat X_m$
is the average maximum of the shower at 1~km from the core  
($\simeq$~200~g cm$^{-2}$ deeper than at the core) 
and $\Lambda\simeq 70$~g cm$^{-2}$ is an  effective hadronic attenuation
length. Then,  under a pressure change, the electromagnetic component $S_{em}$
of $S(1000)$ changes by %%
\begin{equation}
\label{dsdp}
\frac{{\rm d\ ln}S_{em}}{{\rm d}P}=
-\left[1-\frac{\hat X_m}{X}\right]\frac{{\rm sec}\,\theta}{\Lambda},
\end{equation}
where d$X={\rm d}P\,{\rm sec}\,\theta$ was used. Since for the energies of
interest, $E>10^{18}$~eV, the maximum of vertical showers is close to ground,
this effect is expected to be more pronounced for inclined showers.\\
%Second effect
(ii) an increase in the air density reduces the Moli\`ere radius $r_M$
(proportional to $1/\rho$) and hence the lateral extent  of the electromagnetic
component of the shower. The lateral distribution of the electromagnetic
component can be approximately described with an NKG profile, which for large
radius $r$ from the core behaves as 
$N_{em}(r)\propto r_M^{-2}(r/r_M)^{-\alpha}$, where $\alpha\simeq 4$ and
$r_M\simeq 83\ {\rm m}/(\rho/\ {\rm kg\ m}^{-3})$.
Hence, under a density change
\begin{equation}
\frac{{\rm d\ ln}S_{em}}{{\rm d}\rho}\simeq \frac{(2-\alpha)}{\rho}.
\end{equation}
 The effective value of $r_M$ is that corresponding to the air density
$\rho^{*}$  two cascade units above ground \cite{greisen:1956} ($\sim 700\ {\rm
m}\cos\theta$ at the Auger site, with $\theta$ being the zenith angle). Since
the ground $T$, $P$ are the only available observables, we have to express 
$\rho^*$ in terms of the density $\rho$ measured at ground.

On time scales of one day or more, the temperature gradient in the lowest
layers of the atmosphere (the planetary boundary layer) can be described by  an
average value of  $6.5^\circ$C\ km$^{-1}$; therefore the variation of $\rho^*$
is the same as that of $\rho$. An additional effect is related to the diurnal
variations of the gradient that  is smaller before sunrise, at which time even
$T$ inversions are common, and larger in the early afternoon hours.  As a
result, the amplitude of diurnal variations in $T$ (and $\rho$) is smaller at 2
cu than at ground level by a factor $\simeq 0.5$. 
We define the average daily densities $\rho_{d}$ and $\rho^*_{d}$ and the
reference values (averaged over 2 years of measurements)
$\rho_0=1.055$~kg~m$^{-3}$ and $P_0=861.9$~hPa, $\rho_0^*$ denotes the
reference density at 2 cu above ground.
The energy reconstructed with no correction for weather effects is  $E_{r}
\propto \left[S(1000)\right]^B$, where $B=1.13\pm0.02$  \cite{roth:icrc07}. 
Hence we can parameterise the relation between the  shower energy 
$E_{0}(\theta,P,\rho)$ at the reference weather conditions  and the
reconstructed one $E_r$ as: 
\begin{eqnarray}
\label{eq:energy}
\nonumber  E_{0} & = & E_{r} \{1-\alpha_P(P-P_0)-\alpha_\rho(\rho^*-\rho^*_0)\}^B \\ 
\nonumber        & = & E_{r} \{1-\alpha_P(P-P_0) - \\
                 &   &\alpha_\rho(\rho_{d}-\rho_0)-\beta_\rho(\rho-\rho_{d})\}^B   
\end{eqnarray}
where the coefficients $\alpha_{\rho,P}$ and $\beta_\rho$ depend on the zenith
angle $\theta$.\\
Assuming that the cosmic ray spectrum is a pure power law d$J/{\rm d}E\propto
E^{-\gamma}$, it is easy to show that the rate $R(\theta,P,\rho)$ of events at
a given zenith angle  $\theta$  can be expressed as:
\begin{eqnarray}
\label{eq:fitfunc}
\nonumber   R & = & R_0 \{1+a_P(P-P_0)+\\  
                      &   & a_\rho(\rho_{d}-\rho_0) +b_\rho(\rho-\rho_{d})\}
\end{eqnarray}
with  $R_0 = R(\theta,P_0,\rho_0)$ and coefficients
$a_{\rho,P}=(B\gamma-1)\alpha_{\rho,P}$ and $b_\rho = (B\gamma-1)\beta_{\rho}$,
the latter  describing the diurnal modulation of the rate with the density.
\section{Modulation of the measured rates of events}
To study the modulation of the event rate with the ground weather parameters,
we  use the data taken from 1 January 2005 to 31 December 2006 that  have a
zenith angle $\theta < 60^\circ$.  The data selection criterion is the same as
applied for the SD spectrum \cite{roth:icrc07}. The value of the air density
$\rho$ at ground  is deduced from $P$ and $T$ measured at the central
meteorological station. 
Rather than using the raw number of triggering events, we compute the rates, as
a function of time, to account for the temporal variation of the active
detection area due mainly to the deployment of new stations and occasionally to
stations experiencing a temporary failure \cite{suomijarvi:icrc07}. The
modulation of the rate  during the year, and as a function of the hour of the
day,  follows the changes in density and pressure (Figs.~\ref{fig:daily} and
\ref{fig:diurnal}). 
 A characteristic of the Malarg\"{u}e site is the stability of pressure (less
than $\pm$2\% variation), while $\rho_{d}$  changes up to a maximum of $\pm$6\%
during the year with an additional diurnal variation of density of $\pm$2\% on
average, with maximum values of $^{+6}_{-8}$\% during the two years considered.
Assuming that the rates computed each hour follow a Poisson distribution, a
maximum likelihood fit  gives the estimated values of the coefficients in
eq.~(\ref{eq:fitfunc})  averaged over the event distribution in the zenith
range $0^{\circ}-60^{\circ}$ :
\begin{eqnarray}
\label{eq:fitvalues}
\nonumber   a_P & = & (-0.0009 \pm 0.0005) \rm{~hPa^{-1}}   \\ 
            a_{\rho} & = & (-2.68 \pm 0.07)~\rm{kg^{-1}~{m}^3}     \\
\nonumber   b_{\rho} & = & (-0.85 \pm 0.07) \rm{~kg^{-1}~{m}^3}  .
\end{eqnarray}
\vspace{-0.5cm}
%%
%%%%%%%%%%%%%%%%%%%%%%%%%%%%%%%%%%%%%%%%%%%%%%%%%%
\begin{figure}[ht]
   \begin{center}
   \vspace{-0.2cm}
   \includegraphics[width=0.48\textwidth]{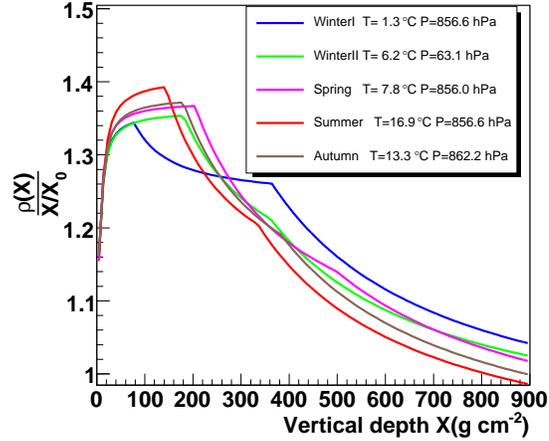}
   \end{center}
   \vspace{-0.4cm}
  \caption{ Seasonal atmospheric profiles derived from the parameterisation of radio soundings performed in Malarg\"{u}e and 
   used in simulations. The density profiles are divided by the profile of 
   an isothermal atmosphere (with $X_{0}$~=~900 g~cm$^{-2}$) to enhance the differences. 
   The corresponding values of P and T are
   given in the box.} 
   \label{fig:atmospheres}
\end{figure}
%%%%%%%%%%%%%%%%%%%%%%%%%%%%%%%%%%%%%%%%%%%%%%%%%%
%
%
%
\section{Comparison of the experimental results with model and simulations }
\begin{figure}[ht]
   \begin{center}
   \includegraphics[width=0.48\textwidth]{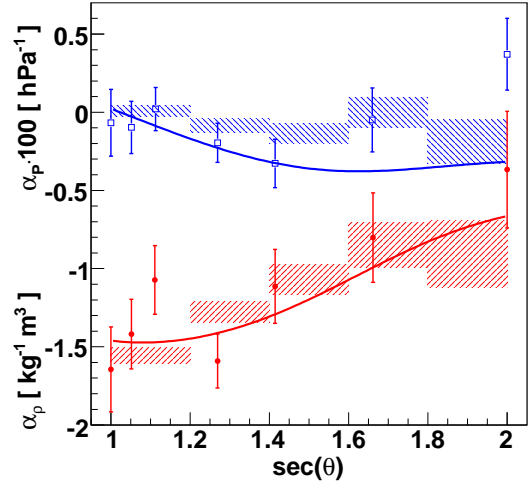}
   \end{center}
   \caption{\label {fig4} Comparison of the signal coefficients $\alpha_{\rho}$ (squares) 
    and $\alpha_{P}$ (circles) obtained from the fit of simulated signal for $10^{19}$ eV proton showers,
    fit of the measured rates (shaded
    rectangles), and 
    values obtained with the theoretical model described in the text (lines). The value of 
    $X_m$~=~750~g~cm$^{-2}$, used in the model,  
     corresponds to 10~EeV showers
    according to the measured elongation rate \cite{unger:icrc07}.} 
    \label{fig:comparison}
\end{figure}

To test the validity of our interpretation, we compare the coefficients 
obtained from the fit of data with results from full shower simulations  and
the predictions on theoretical grounds.\\ 
The Corsika code \cite{heck:corsika98} with the QGSjetII model
\cite{ostapchenko:qgsjet06} for high energy hadronic interactions, was used to
simulate a set of proton showers at $10^{19}$ eV in 5 different atmospheres and
at various zenith angles. 
 The atmospheric profiles used (Fig.~\ref{fig:atmospheres}) are a
parameterisation of the seasonal averages of several radio soundings carried
out at the detector site \cite{keilhauer:2004} and provide a sample of
realistic conditions above the Auger SD array, but, being averages on large
time scales, do not account for the diurnal variation  of the temperature in
the lower atmosphere.
 The expected signal $S(1000)$ is estimated through the simplified assumptions
that $e^{+}$, $e^{-}$ and photons deposit all their energy in the surface
detector, while for muons we take the minimum between the kinetic energy and
240 MeV (the energy deposited  by a vertical muon crossing a SD tank). As
expected, the simulated signal depends on the ground density and pressure
according to the expression in eq.~(\ref{eq:energy}) (with  $\rho-\rho_{d}=0$)
with coefficients $\alpha_{\rho}$ and  $\alpha_{P}$ shown in
Fig.~\ref{fig:comparison} for all zenith angles between 0$^\circ$ and
60$^\circ$. The large uncertainties are due to the limited number of
atmospheric profiles used.\\
For the theoretical expectations, we consider the variation of the total
signal, given by the sum of the electromagnetic and muonic component. The
coefficients $\alpha_{\rho,P}$ in (\ref{eq:energy}) result from the variation
of both  components:  $\alpha_{\rho,P} =
F_{em}\alpha_{\rho,P}^{em}+(1-F_{em})\alpha_{\rho,P}^{\mu}$,  where $F_{em}$ is
the electromagnetic fraction at 1~km.
The dependence of $S_{em}$ on $\rho$ and $P$ is discussed in section 2. For a
quantitative prediction  we adopt in eq.~(\ref{dsdp}) $\hat
X_m=950$~g~cm$^{-2}$, typical of 10~EeV proton showers, and  $X=880\,\rm
sec\,\theta $~g~cm$^{-2}$.  For the electromagnetic fraction $F_{em}$ we use a
fit to the results of shower simulations with 10~EeV protons ($F_{em}\simeq
0.7$ near the vertical and decreasing with $\theta$ to reach $\sim 0.2$ at $60
^\circ$). 
We assume a negligible correlation of $S_{\mu}$ with pressure and a
constant value $\alpha^{\mu}_\rho=-0.26$~kg$^{-1}$~m$^{3}$ for the dependence
on density (suggested by the results of simulations). 
In Fig.~\ref{fig:comparison} we compare the coefficients obtained by fitting
the data in five zenith ranges.  The procedure to obtain $a_{\rho,P}$ is the
same described in section 3, then we derive the signal coefficients  
$\alpha_{\rho,P}$ dividing by $(B\gamma -1) = 2$. Their values are in good
agreement  with both the model predictions and the results from
simulations.    
\section{Conclusions}
The modulation of the event rates measured by the Auger SD can be explained by
known effects on the shower development, both on seasonal and diurnal scales.
At the Auger site the dominant effect is related to the density (temperature)
variation. The systematic error, when determining the  energy of a single
shower in the zenith range  $0^\circ-60^\circ$, amounts to a maximum of
$\sim10\%$  (for extreme values of ground pressure and temperature). The
quantitative agreement of the  theoretical model with simulations and data,
suggests that it can be used to correct the SD energy  reconstruction for
weather induced effects.
%%%%%%%%%%%%%%%%%%%%%%%%%%%%%%%%%%%%%%%%%%%%%%%%%%
%%%%%%%%%%%%%%%%%%%%%%%%%%%%%%%%%%%%%%%%%%%%%%%%%%
%
%
%%%%%%%%%%%%%%%%%%%%%%%%%%%%%%%%%%%%%%%%%%%%%%%%%%%

%This is the reference to .bib file (Without .bib!)
\bibliography{icrc0302}

%This in the bibtex style, is ok.
\bibliographystyle{amsplain}

\end{document}